\documentclass[useamsfonts]{pasj02}

\jyear{2024}
\Received{2024/04/05}
\Accepted{2024/12/04}
\usepackage{anyfontsize}

\usepackage{bm}
\usepackage{xspace}

\newcommand{\Obsv}{\tilde{\bm{v}}}
\newcommand\Comp{\mathbb{C}}
\newcommand{\argmin}[1]{\mathop{\mathrm{arg\,min}}_{#1}}
\newcommand\uv{$u$--$v$\xspace}
\newcommand{\EstS}{Small variance}
\newcommand{\EstL}{Large variance}

\graphicspath{{figures}{.}}

\begin{document}

\title{Solving Self-calibration of ALMA Data with an Optimization Method}

\author{Shiro \textsc{Ikeda}\altaffilmark{1,2,3,4,$\ast$}\orcid{0000-0002-2462-1448}
\email{shiro@ism.ac.jp}
{Takeshi \textsc{Nakazato}\altaffilmark{3}}\orcid{0000-0003-3780-8890}
{Takashi \textsc{Tsukagoshi}\altaffilmark{5}}\orcid{0000-0002-6034-2892}
{Tsutomu T.\ \textsc{Takeuchi}\altaffilmark{6,1}}\orcid{0000-0001-8416-7673}
{Masayuki \textsc{Yamaguchi}\altaffilmark{7}}}\orcid{0000-0002-8185-9882}

\altaffiltext{1}{The Institute of Statistical Mathematics, 10-3 Midori-cho, Tachikawa, Tokyo 190-8562, Japan}
\altaffiltext{2}{Department of Statistical Science, School of Multidisciplinary Sciences, Graduate University for Advanced Studies (SOKENDAI),
10-3 Midori-cho, Tachikawa, Tokyo 190-8562, Japan}
\altaffiltext{3}{National Astronomical Observatory of Japan, 2-21-1 Osawa, Mitaka, Tokyo 181-8588, Japan}
\altaffiltext{4}{Kavli IPMU, The University of Tokyo, 5-1-5 Kashiwanoha, Kashiwa, Chiba 277-8583, Japan}
\altaffiltext{5}{Division of Systems and Information Engineering, Ashikaga University, 268-1 Omae-cho, Ashikagashi, Tochigi 326-8558, Japan}
\altaffiltext{6}{Division of Particle and Astrophysical Science, Nagoya University, Furo-cho, Chikusa-ku, Nagoya, Aichi 464-8602, Japan}
\altaffiltext{7}{Academia Sinica Institute of Astronomy and Astrophysics, 11F of ASMA Building, No.1, Sec. 4, Roosevelt Rd, Taipei 106216, Taiwan}


\maketitle

\begin{abstract}
We reformulate the gain correction problem of the radio interferometry as an optimization problem with regularization, which is solved efficiently with an iterative algorithm. Combining this new method with our previously proposed imaging method, PRIISM, the whole process of the self-calibration of radio interferometry is redefined as a single optimization problem with regularization. As a result, the gains are corrected, and an image is estimated. We tested the new approach with ALMA observation data and found it provides promising results. 
\end{abstract}

\section{Introduction}
\label{sec:intro}

Radio interferometry is one of the important observation methods in astronomy to obtain high-resolution images of celestial objects. For example, the Atacama Large Millimeter/sub-millimeter Array (ALMA) telescope, a world-leading radio interferometer, has significantly contributed to astronomy and astrophysics. We also remember the images of supermassive black holes, M87 and SgrA*, were obtained by a Very Long Baseline Interferometer (VLBI), the Event Horizon Telescope (EHT), which is another example of radio interferometry.

In radio interferometry, the information corresponding to the 2-D Fourier transform of the image, known as visibility, is obtained by combining observations from multiple stations. Although the image is computed by applying the inverse Fourier transform to the visibility in principle, solving this inverse problem is difficult because observations are noisy and the information in the Fourier domain is not sampled uniformly. Traditionally, the CLEAN method has long been used in radio interferometry imaging \citep{1974.Hogbom.CLEAN,1980.Clark.CLEAN,1984.Schwab.CLEAN}. This is equivalent to the Matching Pursuit (MP) method \citep{1993.MallatZhang.ieeesp}, which estimates a collection of points in the image domain to explain the visibility through a greedy process \citep{2021.Ellien_eatl.aa}. With the developments of signal processing theories, the problem of the MP method is reformulated as an optimization problem with regularization, and modern signal processing algorithms have been applied to the imaging of radio interferometry \citep{2009.Wiaux.mnras, 2014.Honma.pasj}. The EHT collaboration proposed related methods for VLBI and named them the Regularized Maximum Likelihood (RML) methods, which contributed to the images of the black hole shadows \citep{eht-imaging, SMILI}. We also have developed and released the RML software Python module for Radio Interferometry Imaging with Sparse Modeling (PRIISM) for imaging with ALMA \citep{2020.NakazatoIkeda.priism}.

In this paper, we focus on the calibration of visibility. The gain of each telescope varies over time, leading to changes in the phase and magnitude of visibility. The calibration is the process of estimating and canceling the influence of the gain variations. This is an important process to increase the quality of the reconstructed images. 

A widely used method for estimating gains is self-calibration, where the gains are adjusted based on the reconstructed image. By iterating image estimation and gain correction, an image is reconstructed. This is a powerful strategy, but it requires careful treatment. In this paper, we reformulate the gain correction as an optimization problem with regularization and propose an efficient iterative algorithm to solve it. With the new formulation, the whole process of gain correction and imaging is reformulated as a single optimization problem. We tested the method with ALMA data and obtained promising results.

The rest of the article is organized as follows: Section 2 presents the theoretical background, Section 3 introduces the proposed method, and Section 4 shows the imaging results with ALMA data before the conclusion in Section 5.

\section{Radio Interferometry}
\label{sec:radio_interferometry}

\subsection{Visibility and Imaging}

After observing a celestial object with multiple radio telescope stations simultaneously, the correlations of the recorded time series are computed for each pair of stations and converted into complex values. The collection of these complex values, $\bm{v}=(v_1,\cdots,v_M)^T\in\Comp^M$ (where $M$ is the number of the data points and $^T$ denotes transpose), referred to as visibility, corresponds to the two-dimensional Fourier transform of the image $\bm{x}$.
Let us denote the ideal Fourier transform relation between visibility $\bm{v}$ and image $\bm{x}$ as follows,
\begin{equation}
    \label{eq:ideal}
    \bm{v} = \mathcal{F}(\bm{x}).
\end{equation}
We focus on total intensity (Stokes I) imaging and assume $\bm{x}$ is ${N_x\times N_y}$ pixelated non-negative image. For observed visibility $\Obsv = (\tilde{v}_1,\cdots,\tilde{v}_M)^T$, the Equation (\ref{eq:ideal}) holds approximately, that is,
\begin{equation}
    \label{eq:Fourier}
    \Obsv \simeq \mathcal{F}(\bm{x}).
\end{equation}
This is due to the additive thermal noise and fluctuations in each station's gain. Initially, gains of the antennas and the noises are not well constrained. The gains are estimated, and the visibilities are calibrated in one set of processes. For most modern arrays, initial calibration measurements are made instrumentally or from astrophysical sources with known properties and applied to the target. However, this initial calibration cannot remove the gain fluctuations that happen during the observation due to the atmospheric conditions. We need further correction of the visibilities.

Each $\tilde{v}_k$ is associated with the observation time for integration, indexed by $t_l$, and the ordered pair of stations, such as stations $\alpha$ and $\beta$, and there is a one-to-one correspondence between the index of the visibility $k$ and the triplets.
\begin{equation}
    \label{eq:relation of k and gains}
    k \rightarrow (t_l,\alpha,\beta).
\end{equation}
The visibility is influenced by the antenna response. We assume that the antenna response can be effectively adjusted with a single complex coefficient for each integration. Let us define the gain correction coefficient of the station $\alpha$ at time $t_l$ as $g_{\alpha l}\in\Comp$. With an abuse of terminology, we refer to this coefficient as ``gain'' in this paper, and $\bm{g}$ denotes the collection of gains. 

Following the above discussion, the observation equation corresponding to equation\,(\ref{eq:Fourier}) is defined as follows,
\begin{equation}
    \label{eq:observation equation}
    \tilde{v}_k g_{\alpha l} g^{\ast}_{\beta l} = \mathcal{F}_k(\bm{x})+n_k,
\end{equation}
where $^\ast$ and $\mathcal{F}_k$ denote the complex conjugate, the Fourier transform for the component $k$, and $n_k$ is the noise, respectively. We define $\bm{n} = (n_1,\cdots,n_M)^T$ as a noise vector. It is important to adjust $\bm{g}$ in order to have a good image.

The image $\bm{x}$ is estimated from $\Obsv$ without directly observing $\bm{g}$ or $\bm{n}$. This ``estimation process'' is separated into two parts: estimating the gains $\bm{g}$, referred to as gain correction, and estimating the image $\bm{x}$ from the calibrated visibility. We explain each part, starting with the latter one.

\subsection{Estimating an image from calibrated visibility}

Let us define calibrated visibility as
\begin{equation}
    \label{eq:gains}
    \hat{\bm{v}} = (\hat{v}_1,\cdots,\hat{v}_M)^T,\hspace{2em}
    \hat{v}_k = \tilde{v}_k \hat{g}_{\alpha l}\hat{g}^{\ast}_{\beta l},
\end{equation}
where $\hat{g}_{\alpha l}$ represents the estimated gain.
Equation (\ref{eq:observation equation}) is rewritten as
\begin{equation}
    \label{eq:calibrated observation equation}
    \hat{\bm{v}} = \mathcal{F}(\bm{x})+\bm{n}.
\end{equation}
Estimating $\bm{x}$ from the observation represented with equation\,(\ref{eq:calibrated observation equation}) is equivalent to performing the inverse Fourier transform from noisy observation. A common approach is to consider the following minimization problem. This is equivalent to the Maximum Likelihood Estimation.
\begin{equation}
\label{eq:CLEAN}
\hat{\bm{x}}=
\argmin{\bm{x}}
  L_{\Obsv}(\bm{x},\hat{\bm{g}})
    \hspace{1em}
    {\mbox{subject to}}
    \hspace{1em}
    \bm{x}\ge 0,
\end{equation}
where
\begin{equation}
    \label{eq:chi_sq}
    L_{\Obsv}(\bm{x},\bm{g})
    =
    \sum_{k} \frac{1}{2\sigma_k^2}\bigl| \tilde{v}_k g_{\alpha l}g^\ast_{\beta l} - \mathcal{F}_k(\bm{x})\bigr|^2.
\end{equation}
Solving the above problem is not easy for radio interferometry because the positions of visibility components are not uniformly distributed in the Fourier domain.

For over forty years, this inverse problem has been solved by the so-called CLEAN method \citep{1974.Hogbom.CLEAN,1980.Clark.CLEAN,1984.Schwab.CLEAN}. It assumes that the image can be built up from point sources and solves the problem through a greedy approach. Recently, the RML approach has been demonstrated as a powerful method for solving the problem \citep{2014.Honma.pasj,2018.Kuramochi.apj,2018.Chael.aj}. The EHT used both CLEAN and RML methods \citep{SMILI,eht-imaging} for imaging the black hole shadows \citep{2019.EHT.M87.paper4,2022.EHT.SgrA.paper3}. We would like to emphasize that the RML approach is flexible, as different functions can be incorporated into regularization.

Based on the success of the RML approach, we have developed a Python code for ALMA imaging and released it to the public as PRIISM \citep{2020.NakazatoIkeda.priism}. The method has been reported to provide better resolution than CLEAN for ALMA data \citep{2020.Yamaguchi.apj,2021.Yamaguchi.Apj,2024.Yamaguchi_etal.pasj}. We will explain the details in \S\,\ref{subsec:priism}.

\subsection{Gain correction}

Atmospheric conditions continuously change, and the prior corrections (such as from phase referencing) do not remove target gain errors on short timescales and/or are not completely accurate for the target direction. The purpose of estimating gains is to reduce the effects of these fluctuations and improve the image accuracy. This is more difficult compared to estimating an image from calibrated visibility. The method of self-calibration derives corrections of gains using the target data themselves (See next subsection).

We describe the gain correction, which is the process to estimate $\bm{g}$, by an expression that minimizes $L_{\Obsv}(\bm{x},\bm{g})$ (Equation (\ref{eq:chi_sq})) w.r.t. $\bm{g}$ for a given $\hat{\bm{x}}$, that is,
\begin{equation}
\label{eq:selfcal}
\hat{\bm{g}}=
    \argmin{\bm{g}}
  L_{\Obsv}(\hat{\bm{x}},\bm{g}).
\end{equation}
Generally, the number of gains is smaller than $M$ because gains are associated with each antenna, while visibility components are computed from pairs of antennas. However, the above problem is non-convex, indicating it has local minima. In order to avoid them, the problem is solved under different constraints, and the gain correction is more common to be human-supervised than to be fully automated. In this paper, we propose a new method for this problem. The details will be explained in \S\,\ref{subsec:new self calibration}.

\subsection{Current approaches to self-calibration\label{subsec:self-calibration}}

The self-calibration (\cite{1980.Schwab.spie}; \cite[Chapter 11.3.2]{2017.ThompsonMoranSwenson.book}) is a method to estimate both an image and gains. It iterates CLEAN and gain correction steps alternately until some condition is satisfied. This powerful method is widely used for radio interferometry imaging \citep{2018.Brogan_etal.arxiv,ALMAmemo620}. 

In this paper, we replace these steps with a new method (see Section \ref{sec:proposed method}). A similar approach has been proposed for the Square Kilometre Array (SKA) \citep{2017.Repetti.mnras}. Here, we note that the RML methods proposed for the EHT have implemented a different approach from the self-calibration \citep{2019.EHT.M87.paper4,SMILI,eht-imaging}. Instead of self-calibration, these methods directly estimate the image by evaluating the errors of closure phases and closure amplitudes. Each closure quantity is defined with more than two visibility components and, theoretically, is not influenced by multiplicative per-antenna gain errors. This is a promising direction for the research, but we stay with the iterative method for ALMA, since the larger number of antennas and easier prior calibration means it is usually easier to obtain images, but on the other hand, the computation of closure quantities is demanding.

\section{Proposed method\label{sec:proposed method}}
\subsection{PRIISM: an RML method for imaging\label{subsec:priism}}

The RML method optimizes the likelihood function under some constraints on the images, which are represented in the regularization. There are different types of regularization, and we follow our proposal in \cite{2018.Kuramochi.apj} where $L^1$ norm and the Total Squared Variation (TSV) for promoting sparsity and smoothness of the image, respectively, were employed. The negative log-likelihood function is defined in equation\,(\ref{eq:chi_sq}), and the optimization problem is defined as follows.
\begin{eqnarray}
    \label{eq:imaging}
    &&
    \hat{\bm{x}}=\argmin{\bm{x}} \Bigl[L_{\Obsv}(\bm{x},\hat{\bm{g}}) + R_{\lambda_1,\lambda_2}(\bm{x})\Bigr]
    \\
    \nonumber
    &&\mbox{subject to}
    \hspace{1em}
    \bm{x}\ge 0,
\end{eqnarray}
where,
\begin{eqnarray}
    &&R_{\lambda_1,\lambda_2}(\bm{x})=
    \lambda_1\|\bm{x}\|_1 + \lambda_2\mathrm{TSV}(\bm{x}),\hspace{0.5em}\lambda_1,\lambda_2\ge 0, \\
    \nonumber
    &&\|\bm{x}\|_1= \sum_{ij}|x_{ij}|,\\\nonumber
    &&\mathrm{TSV}(\bm{x}) = \sum_{ij}(x_{ij} - x_{i-1\,j})^2 +(x_{ij} - x_{i\,j-1})^2.
\end{eqnarray}
Here, $\| \cdot \|_1$ and $\mbox{TSV}(\bm{x})$ are the $L^1$ norm, and the TSV of $\bm{x}$, respectively. Once positive coefficients $\lambda_1$ and $\lambda_2$ are given, this optimization problem is convex with respect to $\bm{x}$. We have implemented the Monotonic Fast Iterative Shrinkage-Thresholding Algorithm (MFISTA) introduced in \cite{2009.BeckTeboulle.siam}, which is the accelerated version of the proximal gradient descent method to solve this problem. For the computation of $\mathcal{F}_k(\bm{x})$, we realized the NonUniform FFT (NUFFT) described in \cite{2004.LeslieLee.siam} with the FFTW3 package. Our implementation is available as PRIISM on GitHub \citep{2020.NakazatoIkeda.priism}.

\subsection{Regularized optimization problem for gain correction\label{subsec:new self calibration}}

We extend the idea in \cite{2017.Repetti.mnras} and build a self-calibration algorithm for a smaller field of view, such as that of ALMA, assuming that the gain of each telescope does not change within the field of view but changes over time. We further assume that the gains of telescopes change smoothly over time. The gain correction problem is formulated as follows.
\begin{eqnarray}
\label{eq:selfcal_new}
&&\hat{\bm{g}}
=
\argmin{\bm{g}}
\Bigl[L_{\Obsv}(\hat{\bm{x}},\bm{g}) + S_{\mu_1,\mu_2}(\bm{g})\Bigr]\\\nonumber
&&\mbox{subject to}
    \hspace{1em}
 \sum_{l} |g_{\alpha l}| = 
   N_\alpha,
\end{eqnarray}
where, $N_{\alpha}$ is the number of gains of station $\alpha$ and 
\begin{eqnarray}
\label{eq:gain_regularisers}
&&
S_{\mu_1,\mu_2}(\bm{g}) =
\mu_1
\sum_{\alpha,l}
 {w_{\alpha l}}
 \bigl| g_{\alpha l}-g_{\alpha\,l-1}\bigr|^2
\\\nonumber
&&\hspace{6em}+\mu_2
\sum_{\alpha,l}
 {w_{\alpha l}}
   \bigl(|g_{\alpha l}|-|g_{\alpha\,l-1}|\bigr)^2,\\\nonumber
 &&\mu_1, \mu_2 \ge 0, \hspace{1em} w_{\alpha l} = \frac{1}{t_l - t_{l-1}}.
\end{eqnarray}
We will discuss how to set positive coefficients $\mu_1$ and $\mu_2$ in \S\,\ref{subsec:parameters}. The first term of $S_{\mu_1,\mu_2}(\bm{g})$ penalizes differences in both the amplitude and phase of gains between adjacent times. The second term only considers the amplitude. The weight $w_{\alpha l}$ controls the similarity between gains adjacent in time. If the adjacent times are close, $w_{\alpha l}$ becomes large and enforces the gains to be close. The constraint of the gains in equation\,(\ref{eq:selfcal_new}) prevents the gain from shrinking to $0$. Note that $N_\alpha$ may differ from each station because not all stations contribute to the observations at every instance.

The objective function in equation\,(\ref{eq:selfcal_new}) is quartic and non-convex w.r.t. $\bm{g}$. We followed the idea in \cite{2017.Repetti.mnras} for the optimization, which is described in Appendix \ref{sec:details of algorithm}. The estimated gains can become $0$. If it happens, the corresponding visibility component is flagged. We note that it did not happen in the experiments shown in section \ref{sec:examples}.

\subsection{Overall imaging\label{subsec:overall imaging}}

To estimate both the image $\bm{x}$ and the gain $\bm{g}$, equations\,(\ref{eq:imaging}) and (\ref{eq:selfcal_new}) are iterated as follows.
\begin{enumerate}
    \item Initialize
    the gain as $\hat{\bm{g}} =\bm{1}$.
    \item Set $\bm{g}$ to $\hat{\bm{g}}$ and update the visibilities with equation\,(\ref{eq:gains}) using $\hat{\bm{g}}$. With the updated visibilities, estimate ${\bm{x}}$ by solving equation\,(\ref{eq:imaging}).
    \item Set $\bm{x}$ to $\hat{\bm{x}}$ and estimate ${\bm{g}}$ by solving equation\,(\ref{eq:selfcal_new}).
    \item Iterate steps 2 and 3 until it converges or the maximum number of iterations is reached. 
\end{enumerate}

In the following examples, we set the maximum iterations of steps 2 and 3 to 10. 

The estimation of the image $\bm{x}$ and the gain $\bm{g}$ are summarised in a single optimization problem by combining equations\,(\ref{eq:imaging}) and (\ref{eq:selfcal_new}) as follows,
\begin{eqnarray}
    \label{eq:overall}
    &&{\hat{\bm{x}},\hat{\bm{g}}} = \argmin{\bm{x},\bm{g}} 
    \Bigl[
       L_{\Obsv}(\bm{x},\bm{g}) + R_{\lambda_1,\lambda_2}(\bm{x}) + S_{\mu_1,\mu_2}(\bm{g})
    \Bigr]\\\nonumber
    &&\mbox{subject to}\hspace{1em} \bm{x}\ge 0, \hspace{.5em}\sum_l | g_{\alpha l}| = N_{\alpha}.
\end{eqnarray}
This is convex with respect to $\bm{x}$ when $\bm{g}$ is fixed, but the overall problem is non-convex. Our approach is to solve it with $\bm{x}$
 and $\bm{g}$ alternately. Generally, it is difficult to find a global optimum of a non-convex optimization problem, and it applies to our approach.

Reformulating the imaging of the radio interferometry as an optimization problem in equation\,(\ref{eq:overall}) allows us to incorporate different types of regularization both for imaging and for self-calibration. We also note that it is possible to apply different optimization algorithms to solve the problem defined in equation\,(\ref{eq:overall}) \citep{2023.Takahashi.phd}.

\subsection{Choosing the parameters \label{subsec:parameters}}

The imaging problem described in equation\,(\ref{eq:overall}) has four parameters $\lambda_1$, $\lambda_2$, $\mu_1$ and $\mu_2$. Setting these parameters to appropriate values is crucial since the final image depends on them. A natural approach is to define a statistical measure and determine the best parameters by testing different combinations. Finding a good statistical measure and efficiently testing various combinations pose significant challenges. In this paper, we present one method for parameter selection, but there will be alternative methods. Studying them will be one of the future tasks.

The parameters $\{\lambda_1,\lambda_2\}$ control the regularization of $\bm{x}$, where larger $\lambda_1$ and $\lambda_2$ promote the sparsity and the smoothness of the image, respectively. The parameters $\{\mu_1,\mu_2\}$ control regularization of gains, where larger $\mu$'s suppress the variation of the gains in time: large $\mu_1$ penalizes the variations of amplitude and phase, while large $\mu_2$ penalizes the variations of amplitude of the gains. Because these two sets work differently, we defined separate statistical measures for $\{\lambda_1, \lambda_2\}$ and $\{\mu_1, \mu_2\}$. We explain them before showing how we explore different combinations efficiently. In the following, we consider ALMA data, where the number of \uv points is large and telescopes have similar gains.

\begin{figure*}
    \begin{center}
        \includegraphics[width=\textwidth]{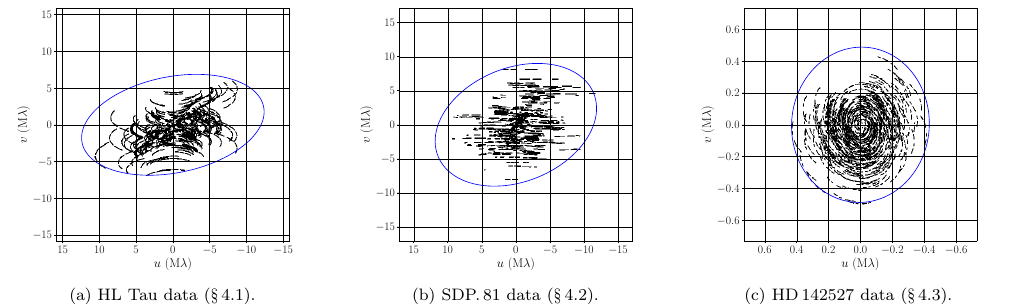}
    \end{center}
    \caption{The \uv coverages of the data used in \S\,\ref{sec:examples}. Black points are the observed \uv points, and the blue lines show the covering \uv ellipsoids.
    Alt text: Three scatter plots. Each plot is accompanied by an ellipsoid, which includes all the points inside.}
    \label{fig:u-v ellipsoid}
\end{figure*}

Firstly, we define the statistical measures for choosing $\lambda_1$ and $\lambda_2$. The weighted squared error $L_{\Obsv}(\bm{x},\bm{g})$ is a statistical measure for $\bm{x}$. In our previous works \citep{2018.Nakazato.ADASS,2020.Yamaguchi.apj}, the generalization error of $L_{\Obsv}(\bm{x},\bm{g})$ was evaluated through the Cross-Validation (CV) method in order to avoid the risk of overfitting, and was used for the selection of $\lambda_1$ and $\lambda_2$. This is a standard method in data science, but we use a different approach in this paper because the computation of CV is demanding. Instead, we pose two conditions for $\bm{x}$ and choose $\lambda$'s, which minimizes the $L_{\Obsv}(\bm{x},\bm{g})$ under the conditions.
\begin{enumerate}
    \item Most of the power of the image in the \uv domain is included in the ``covering \uv ellipsoid.''
    \item The weighted squared error of the visibility
    \begin{equation}
    \label{eq:chi_sq_k}
    \frac{1}{2\sigma_k^2}
    \bigl| \tilde{v}_k g_{\alpha l}g^\ast_{\beta l} - \mathcal{F}_k(\bm{x})\bigr|^2
    \end{equation}
    is uniform over the \uv distance.
\end{enumerate}
The first condition constrains the image $\bm{x}$ in the \uv domain. Here, the ``covering \uv ellipsoid'' is the smallest centered ellipsoid, which covers all the \uv points (Figure \ref{fig:u-v ellipsoid}). The image $\bm{x}$ is converted into the \uv domain by the Fourier transform, and the power of the image in the covering \uv ellipsoid is divided by the total power of the image. We define this ratio as $C_1$, and set the condition as $C_1 \ge C_1^\ast$, where $C_1^\ast$ is set to $0.995$ in \S\,\ref{sec:examples}. Roughly speaking, large $\lambda_2$ makes the image smoother and $C_1$ larger.

For the second condition, the visibility components are split into three groups depending on the \uv distance (small, mid, large), and the averages of the weighted squared errors in equation\,(\ref{eq:chi_sq_k}) of the three groups were turned into a single three-dimensional vector. We define the cosine between the vector and a reference vector $(1,1,1)$ as $C_2$ and set the condition as $C_2 \ge C_2^\ast$, where $C_2^\ast$ is set to $0.99$ in \S\,\ref{sec:examples}.

Secondly, we define the statistical measure for $\mu_1$ and $\mu_2$. It is not appropriate to consider the statistical measure based on the image for setting $\mu_1$ and $\mu_2$ because the main cause of the gain fluctuations is the weather condition. For the statistical measure for $\mu_1$ and $\mu_2$, we focus on the standard deviations of the gains' phase $\sigma_{ph}$ and the gains' amplitude as $\sigma_{amp}$. The parameters $\mu_1$ and $\mu_2$ are chosen to make $(\sigma_{ph}, \sigma_{amp})$ close to a given target values $(\sigma_{ph}^\ast, \sigma_{amp}^\ast)$. 
We set the target values by hand in this work. These values are related to the weather conditions of the ALMA site; ALMA observations provide more information to estimate the values, and how to set the values from ALMA observations is one of our future works. In \S\,\ref{sec:examples}, we tested two sets of $(\sigma_{ph}^\ast, \sigma_{amp}^\ast)$ for each image reconstruction in order to see the influence of these values; one is $(5\mbox{\,deg}, 0.05)$ and the other is $(15\mbox{\,deg}, 0.10)$, in the following, we refer the former estimates as ``\EstS'' and the latter case as ``\EstL.'' As we allow larger variances for gains, larger gain modulation is allowed, and the algorithm tends to make the image sparser and smoother. 

Finally, we explain how we search for the optimal combination of the four parameters. To choose an appropriate set, testing different combinations is inevitable. However, a naive grid search of four parameters is time-consuming, and we separated the parameters into $\{\lambda_1, \lambda_2\}$ and $\{\mu_1, \mu_2\}$ and optimized them alternatively. We used Bayesian optimization \citep{skopt} implemented in an efficient modern machine learning method, to search for an optimal combination of the parameters. For $\{\lambda_1, \lambda_2\}$, the Bayesian optimization problem is defined as
\begin{eqnarray}
    \label{eq:Bayes_lambda}
    \min_{\lambda_1,\lambda_2} 
    \Bigl[
      L_{\Obsv}(\hat{\bm{x}},\hat{\bm{g}}) + \kappa 
      \bigl(
        h_{sq}(C_1^\ast - C_1) + h_{sq}(C_2^\ast - C_2)
      \bigr)
    \Bigr]\\
    \nonumber
    \mbox{where}\hspace{2em} h_{sq}(t) = 
    \left\{
    \begin{array}{ll}
    0 & t < 0 \\
    t^2 & t \ge 0
    \end{array}
    \right.
    ,
\end{eqnarray}
and we set $\kappa = 10^6$ in \S\,\ref{sec:examples}. For $\{\mu_1, \mu_2\}$, it is defined as
\begin{equation}
    \label{eq:Bayes_mu}
    \min_{\mu_1,\mu_2} 
    \biggl[
       \Bigl(\frac{\sigma_{ph}}{\sigma_{ph}^\ast}   - 1\Bigr)^2 
       + 
       \Bigl(\frac{\sigma_{amp}}{\sigma_{amp}^\ast} - 1\Bigr)^2
    \biggr].
\end{equation}
The two conditions of $\lambda$'s are implemented as soft constraints in equation\,(\ref{eq:Bayes_lambda}). 

We expressed parameters as $\lambda_{1,2} = 10^{\Lambda_{1,2}}$, $\mu_{1,2} = 10^{\mathrm{M}_{1,2}}$ and restricted $\Lambda_{1,2}$ and $\mathrm{M}_{1,2}$ to integers of limited ranges. This implementation reduces the search space, which contributes to the efficient search of the parameter set. In this paper, we used a Python implementation, {\tt scikit-optimize} \citep{skopt}, for the Bayesian optimization. The number of trials is fixed to 30 for {\tt scikit-optimize}. This number is set depending on our time and computation facility, and there is a higher chance of being close to the optimum as the number becomes large. The whole procedure to choose the four parameters becomes as follows:
\begin{enumerate}
    \item Find the best combination of $\{\lambda_1, \lambda_2\}$ for the visibility without self-calibration.
    \item Fix $\{\lambda_1, \lambda_2\}$ for those found in step 1 and search the best combination of $\{\mu_1, \mu_2\}$.
    \item Fix $\{\mu_1, \mu_2\}$ for those found in step 2 and search the best combination of $\{\lambda_1, \lambda_2\}$.
\end{enumerate}
In the following experiment, we repeated steps 2 and 3 for 10 iterations because the image does not change largely after a few iterations.

\section{Imaging ALMA data} \label{sec:examples}

\begin{figure*}
    \begin{center}
        \includegraphics[width=\textwidth]{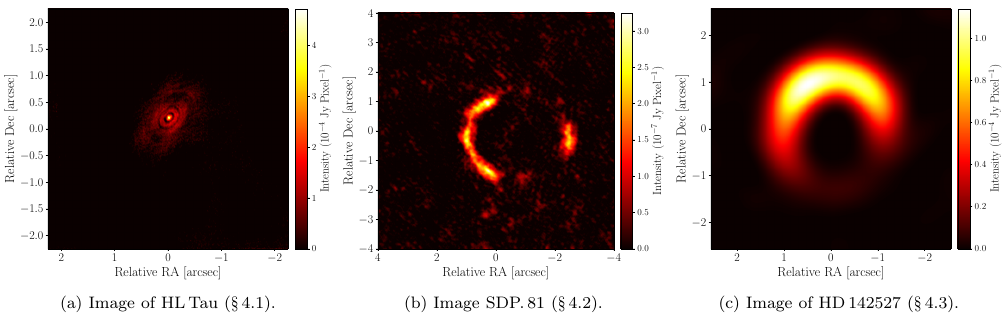}
    \end{center}
    \caption{The estimated images without self-calibration.
    Alt text: Three images.}
    \label{fig:images_uncalibrated}
\end{figure*}
We applied the proposed method to three data sets from ALMA observation: two from ALMA science verification data sets, HL Tau and SDP.\,81, and another data set, HD\,142527. We selected HL Tau and SDP.\,81 from the science verification data sets because they have different characteristics: HL Tau has a distributed structure, while SDP.\,81 consists of a collection of point sources. The third data set, HD\,142527, has been studied in \cite{2020.Yamaguchi.apj}, where the RML method described in 
\S\,\ref{subsec:priism} was applied to the calibrated data. The data size of HD\,142527 is smaller than HL Tau and SDP.\,81. We show the imaging results in the following.

\subsection{HL Tau\label{sec:HLTau}}

\begin{figure*}
    \begin{center}
        \includegraphics[width=\textwidth]{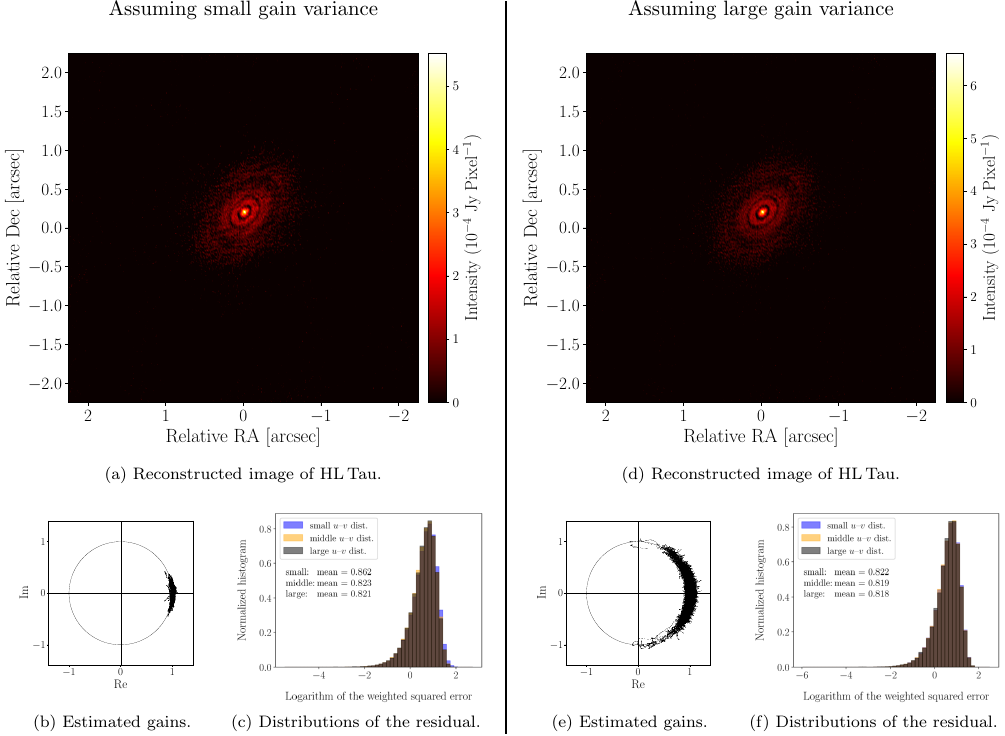}
    \end{center}
    \caption{The imaging results of HL Tau. Smaller variance was assumed for (a), (b), and (c) (\EstS), while larger variance was assumed for (d), (e), and (f) (\EstL):
    (a) and (d) show the reconstructed images.
    (b) and (e) are the scatter plots of the estimated gains in the complex planes.
    We split the \uv points into three groups depending on the equal \uv distances and plot the histograms of the normalized squared residuals in (c) and (f). The black dots in the figures show the mean of the normalized squared distance.
    Alt text: (a): An image. (b): A scatter plot on a complex plane with a unit circle. The points are distributed around $1 + 0 i$ on the arc. (c): Three histograms of different sets of samples are overlapping. The histograms are almost identical. (d): An image. (e): A scatter plot on a complex plane with a unit circle. The points are distributed around $1 + 0 i$ on the arc. The distribution has a wider range than (b). (f): Three histograms of different sets of samples are overlapping. The histograms are almost identical.}
    \label{fig:HLTau_result}
\end{figure*}

The first data set is the HL Tau from the ALMA science verification data \citep{2015.HLTau.apjl}.
This was observed as a part of the 2014 ALMA Long Baseline Campaign \citep{2015.ALMA.apjl}. It consists of observations of 5 days, and we used the data from 29 October 2014 for the imaging. We used a single day of the data to save time for computation. Gains are defined for each integration (minimum averaging interval). The numbers of the \uv points and the gains are 1,884,732 and 116,399, respectively. 

We followed the standard calibration procedure to obtain the calibrated visibility for imaging. The data were calibrated using a tweaked version of the standard calibration scripts provided together with the uncalibrated data. To update the visibility weight value consistently, we changed {\tt calwt} parameter to {\tt True} when bandpass, gain, and flux scaling solutions were applied to the visibility data for HL Tau. We performed the calibration with CASA 4.2.2. The resulting visibility data should be identical to the visibility data for CLEAN-based imaging except for visibility weights.

We used the spectral window 0 out of the total of four spectral windows at Band 6. Because the set of gains is defined for each spectral window, we used only a single spectral window. The size of the image field was set to $4.5\,\mbox{arcsec}\times4.5\,\mbox{arcsec}$, and the number of pixels was set to $512\times512$ pixels.

Figure \ref{fig:images_uncalibrated}a shows the reconstructed image from calibrated visibilities without self-calibration. We applied our method to the same data with two different estimations of gains. The corresponding reconstructed images are shown in figures\,\ref{fig:HLTau_result}a and \ref{fig:HLTau_result}d. The images give sharper impressions, and the peak intensities are higher than that of figure\,\ref{fig:images_uncalibrated}a. The variances of the estimated gains are larger for figure\,\ref{fig:HLTau_result}d than figure\,\ref{fig:HLTau_result}a.

\begin{table}
    \tbl{Selected parameters for imaging for HL Tau. Note that $\Lambda_{1,2} = \log_{10}\lambda_{1,2}$ and $\mathrm{M}_{1,2} = \log_{10}\mu_{1,2}$.}{
    \begin{tabular}{lcccc}
    \hline
        & $\Lambda_1$ & $\Lambda_2$ & $\mathrm{M}_1$ & $\mathrm{M}_2$ \\
        \hline
        \EstS & $-4$ & $8$ & $6$ & $4$ \\
        \EstL & $-4$ & $8$ & $4$ & $4$ \\
        \hline
    \end{tabular}}
    \label{tab:HLTau parameters}
\end{table}

The selected parameters are shown in table\,\ref{tab:HLTau parameters} and the scatter plots of corresponding gains are shown in figures\,\ref{fig:HLTau_result}b and \ref{fig:HLTau_result}e. The gains of figures\,\ref{fig:HLTau_result}b, have larger variances than those of figures\,\ref{fig:HLTau_result}e. From table\,\ref{tab:HLTau parameters} and figures\,\ref{fig:HLTau_result}b and \ref{fig:HLTau_result}e, we can confirm that the variances are controlled by $\mu_1$. Figure\,\ref{fig:HLTau station gains} in Appendix \ref{sec:station gains} shows the phases and the amplitudes of estimated gains as time series. The estimated gains change continuously over time.

\begin{table}
    \tbl{Statistics of the estimated image for HL Tau.}{
    \begin{tabular}{lccccc}
    \hline
    & \multicolumn{2}{c}{Image} & & \multicolumn{2}{c}{Gain} \\
    \cline{2-3} \cline{5-6}
    & $C_1$ & $C_2$ & & $\sigma_{ph}$[deg] & $\sigma_{amp}$ \\ 
    \hline
    {\EstS} & 0.993 & 0.999 & & 4.68 & 0.019 \\
    {\EstL} & 0.994 & 1.000 & & 16.9 & 0.037 \\
    \hline
    \end{tabular}}
    \label{tab:HLTau stats}
\end{table}

Figures\,\ref{fig:HLTau_result}c and \ref{fig:HLTau_result}f show the histograms of the residual amplitudes. The statistics related to the image and the gains are shown in table\,\ref{tab:HLTau stats}. Almost all the power of the image is concentrated in the covering ellipsoid ($C_1 > 0.99$), and the power of the residual is distributed equally over \uv distance ($C_2 > 0.99$). The variances of the estimated gains are not exactly the same as the designed values, but it is difficult to tune the variances with finite trials of the Bayesian optimization. 

\subsection{SDP.\,81\label{sec:SDP81}}

\begin{figure*}
    \begin{center}
        \includegraphics[width=\textwidth]{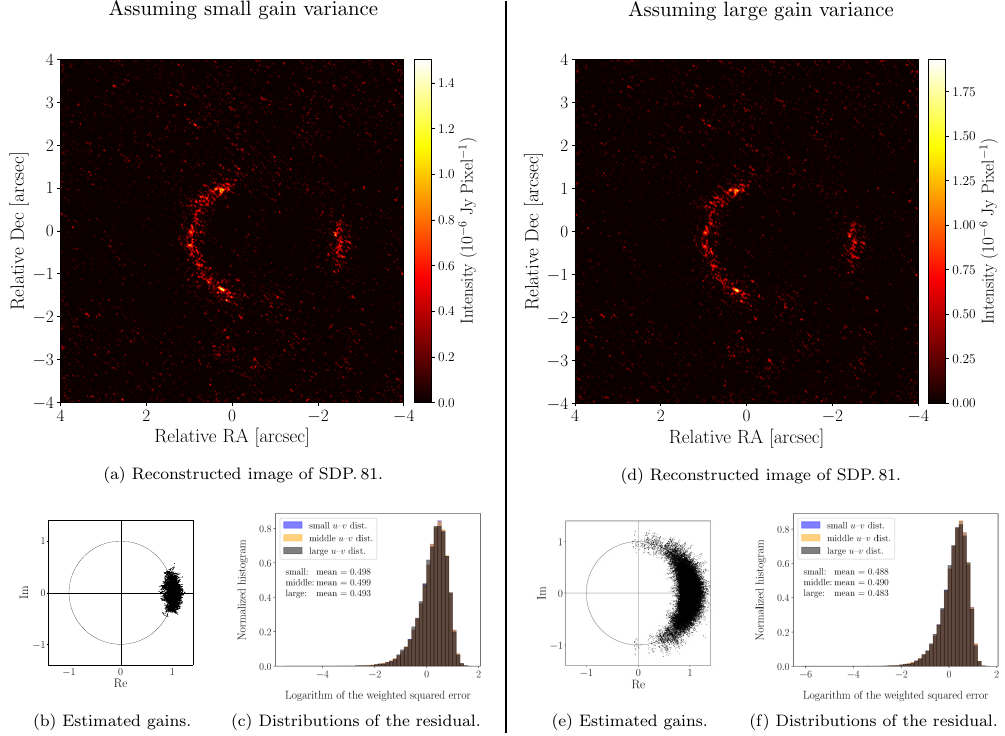}
    \end{center}
    \caption{The imaging results of SDP.\,81. Smaller variance was assumed for (a), (b), and (c) (\EstS), while larger variance was assumed for (d), (e), and (f) (\EstL):
    (a) and (d) show the reconstructed images.
    (b) and (e) are the scatter plots of the estimated gains in the complex planes.
    We split the \uv points into three groups depending on the equal \uv distances and plot the histograms of the normalized squared residuals in (c) and (f). The black dots in the figures show the mean of the normalized squared distance.
    Alt text: (a): An image. (b): A scatter plot on a complex plane with a unit circle. The points are distributed around $1 + 0 i$. (c): Three histograms of different sets of samples are overlapping. The histograms are almost identical. (d): An image. (e): A scatter plot on a complex plane with a unit circle. The points are distributed around $1 + 0 i$. The distribution has a wider range than (b). (f): Three histograms of different sets of samples are overlapping. The histograms are almost identical.}
    \label{fig:SDP81_result}
\end{figure*}

The next data set is the SDP.\,81 from the ALMA science verification data \citep{2015.SDP81.apjl}. This was also observed as a part of the 2014 ALMA Long Baseline Campaign. It consists of observations of 5 days, and we used all of them. Gains are defined for each integration. The number of the \uv points and the gains are 1,650,810 and 110,170, respectively. 

The calibration procedure was exactly the same as HL Tau (\S \ref{sec:HLTau}), i.e., we changed {\tt calwt} parameter for the application of bandpass, gain, and flux scaling solutions to {\tt True} in the standard calibration scripts, and ran tweaked calibration script in CASA 4.2.2.

We used the first spectral channel of Band 6, the size of the image field was set to $8.0\,\mbox{arcsec}\times8.0\,\mbox{arcsec}$, and the number of pixels was set to 2048$\times$2048 pixels.

Figure \ref{fig:images_uncalibrated}b shows the reconstructed image from calibrated visibilities without self-calibration. We applied our method to the same data with two different estimations of gains. The corresponding reconstructed images are shown in figures\,\ref{fig:SDP81_result}a and \ref{fig:SDP81_result}d. The images give sharper impressions, and the peak intensities are higher than that of figure\,\ref{fig:images_uncalibrated}b. The variances of the estimated gains are larger for figure\,\ref{fig:SDP81_result}d than figure\,\ref{fig:SDP81_result}a.

\begin{table}
    \tbl{Selected parameters for imaging SDP.\,81.}{
    \begin{tabular}{l cccc}
    \hline
    & $\Lambda_1$ & $\Lambda_2$ & $\mathrm{M}_1$ & $\mathrm{M}_2$ \\
    \hline
    \EstS  & $-9$ & $12$ & $4$ & $4$ \\
    \EstL & $-9$ & $12$ & $3$ & $1$ \\
    \hline
    \end{tabular}}
    \label{tab:SDP81 parameters}
\end{table}

The selected parameters are shown in table\,\ref{tab:SDP81 parameters} and the scatter plots of corresponding gains are shown in figures\,\ref{fig:SDP81_result}b and \ref{fig:SDP81_result}e. The gains of figures\,\ref{fig:SDP81_result}e, have larger variances than those of figures\,\ref{fig:SDP81_result}b. From table\,\ref{tab:SDP81 parameters} and figures\,\ref{fig:SDP81_result}b and \ref{fig:SDP81_result}e, the variances are controlled by the two parameters $\mu_1$ and $\mu_2$. Figures\, \ref{fig:SDP81 station gains 0} -- \ref{fig:SDP81 station gains 5} in Appendix \ref{sec:station gains} show the phases and the amplitudes of estimated gains as time series. The estimated gains change continuously over time.

\begin{table}
    \tbl{Statistics of the estimated image SDP.\,81.}{
    \begin{tabular}{lccccc}
    \hline
    & \multicolumn{2}{c}{Image} & & \multicolumn{2}{c}{Gain} \\
    \cline{2-3} \cline{5-6}
    & $C_1$ & $C_2$ & & $\sigma_{ph}$[deg] & $\sigma_{amp}$ \\ 
    \hline
    \EstS  & 0.997 & 1.000 & & 5.99 & 0.051 \\
    \EstL & 0.998 & 1.000 & & 16.6 & 0.084 \\
    \hline
    \end{tabular}}
    \label{tab:SDP81 stats}
\end{table}

Figures\,\ref{fig:SDP81_result}c and \ref{fig:SDP81_result}f show the histograms of the amplitudes of residuals. The statistics related to the image and the gains are shown in table\,\ref{tab:SDP81 stats}. Almost all the power of the image is concentrated in the covering ellipsoid ($C_1 > 0.99$), and the power of the residuals is distributed equally over \uv distance ($C_2 > 0.99$). 

\subsection{HD\,142527\label{sec:HD142527}}

\begin{figure*}
    \begin{center}
        \includegraphics[width=\textwidth]{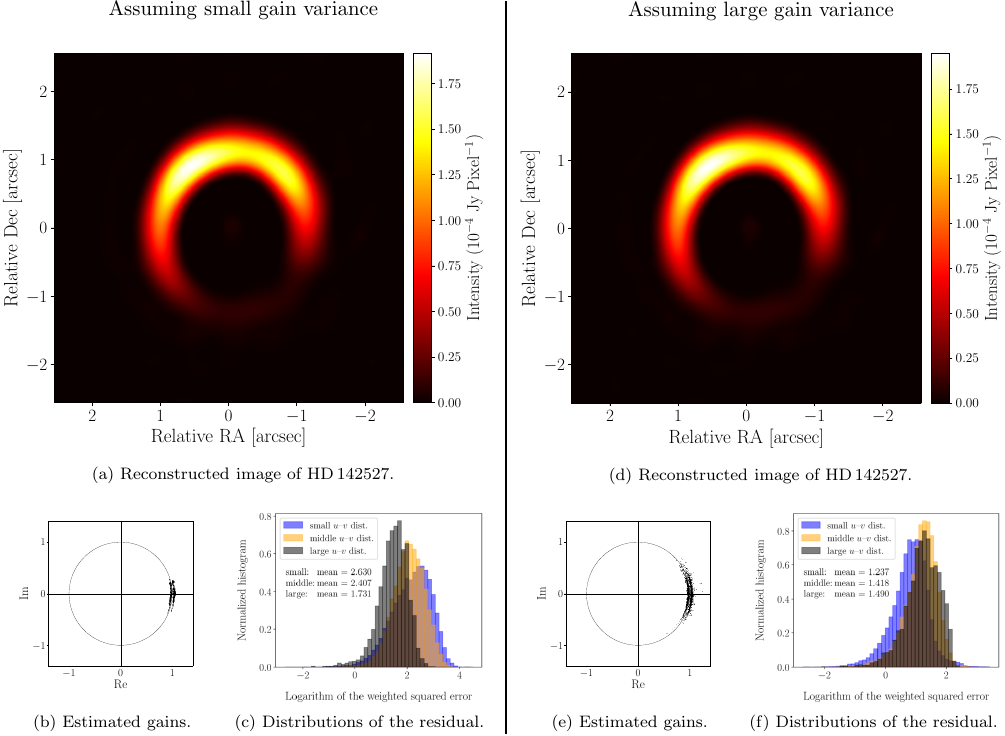}
    \end{center}
    \caption{The imaging results of HD\,142527. Smaller variance was assumed for (a), (b), and (c) (\EstS), while larger variance was assumed for (d), (e), and (f) (\EstL):
    (a) and (d) show the reconstructed images.
    (b) and (e) are the scatter plots of the estimated gains in the complex planes.
    We split the \uv points into three groups depending on the equal \uv distances and plot the histograms of the normalized squared residuals in (c) and (f). The black dots in the figures show the mean of the normalized squared distance.
    Alt text: (a): An image. (b): A scatter plot on a complex plane with a unit circle. The points are distributed around $1 + 0 i$ on the arc. (c): Three histograms of different sets of samples are overlapping. (d): An image. (e): A scatter plot on a complex plane with a unit circle. The points are distributed around $1 + 0 i$ on the arc. The distribution has a wider range than (b). (f): Three histograms of different sets of samples are overlapping.}
    \label{fig:HD142527_result}
\end{figure*}

The last dataset is HD\,142527, which has a protoplanetary disk forming a crescent-shaped ring. The data set is obtained as part of the project 2015.1.00425.S \citep{2016.Kataoka.apjl} and is used for the first time to evaluate ALMA super-resolution imaging with sparse modeling \citep{2020.Yamaguchi.apj}. The data set was calibrated in the same way as \cite{2016.Kataoka.apjl}, using the ALMA pipeline with CASA 4.7.2, except for performing iterative self-calibration. The observations were performed on March 11, 2015, at 343\,GHz (Band 7). The observing array consisted of thirty-eight 12\,m antennas, and the on-source time of HD\,142527 was 1.2\,h. Among the four 2\,GHz spectral windows, the lowest band centered at 336\,GHz was used for image reconstruction.

Figure \ref{fig:u-v ellipsoid} (c) shows the \uv coverage of the data set. Gains are defined for each integration. The numbers of the \uv points and the gains are 49,662 and 3,288, respectively. We set the size of the image field and the number of pixels to $5.12\,\mbox{arcsec}\times5.12\,\mbox{arcsec}$ and $512\times512$ pixels, respectively. The image reconstructed from calibrated visibilities without self-calibration is shown in figure\,\ref{fig:images_uncalibrated}c. We applied our method to the same data with two different estimations of gains. The corresponding reconstructed images are shown in figures\,\ref{fig:HD142527_result}a and \ref{fig:HD142527_result}d. The images give sharper impressions, and the peak intensities are higher than that of figure\,\ref{fig:images_uncalibrated}c. The variances of the estimated gains are larger for figure\,\ref{fig:HD142527_result}d than figure\,\ref{fig:HD142527_result}a.

\begin{table}
    \tbl{Selected parameters for imaging HD\,142527.}{
    \begin{tabular}{l cccc}
    \hline
    & $\Lambda_1$ & $\Lambda_2$ & $\mathrm{M}_1$ & $\mathrm{M}_2$ \\
    \hline
    \EstS & $3$ & $13$ & $9$ & ${-4}$ \\
    \EstL & $2$ & $13$ & $6$ & $4$ \\
    \hline
    \end{tabular}}
    \label{tab:HD142527 parameters}
\end{table}

The selected parameters are shown in table\,\ref{tab:HD142527 parameters} and the scatter plots of the estimated gains are shown in figures\,\ref{fig:HD142527_result}b and \ref{fig:HD142527_result}e. The gains of figures\,\ref{fig:HD142527_result}e, have larger variances than those of figures\,\ref{fig:HD142527_result}b. From table\,\ref{tab:HD142527 parameters} and figures\,\ref{fig:HD142527_result}b and \ref{fig:HD142527_result}e, we can confirm that the variances are controlled by $\mu_1$ and $\mu_2$. Figure\,\ref{fig:HD142527 station gains} shows the phases and the amplitudes of estimated gains as time series. The estimated gains change continuously over time. The estimated gains of each station are plotted in figure\,\ref{fig:HD142527 station gains}. 

\begin{table}
    \tbl{Statistics of the estimated image HD\,142527.}{
    \begin{tabular}{lccccc}
    \hline
    & \multicolumn{2}{c}{Image} & & \multicolumn{2}{c}{Gain} \\
    \cline{2-3} \cline{5-6}
    & $C_1$ & $C_2$ & & $\sigma_{ph}$[deg] & $\sigma_{amp}$ \\ 
    \hline
    \EstS & 0.999 & 0.845 & & 6.04 & 0.031 \\
    \EstL & 0.999 & 0.975 & & 10.79 & 0.034 \\
    \hline
    \end{tabular}}
    \label{tab:HD142527 stats}
\end{table}

Figures\,\ref{fig:HD142527_result}c and \ref{fig:HD142527_result}f show the histograms of the amplitudes of residuals. The statistics related to the image and the gains are shown in table\,\ref{tab:HD142527 stats}. Almost all the power of the image is concentrated in the covering ellipsoid ($C_1 > .99$). The power of the residuals is distributed similarly over \uv distance, but $C_2$ is not larger than 0.9 for \EstS.

\section{Conclusion}

Gain estimation is important for the imaging of radio interferometers. Traditionally, self-calibration has been used, but it requires some hand tuning. In this paper, we have formulated the estimation of gains as an optimization problem. The problem is non-convex, and we propose an efficient method that converges to a local minimum. By alternating the proposed method and RML method iteratively, the image is reconstructed.

We tested the proposed method with three data sets from ALMA observations: HL\,Tau and SDP.81 from ALMA science verification data sets, and HD\,142527. For the estimation of gains, we set their target variances by hand. As larger variances are allowed for the gains, the image becomes sharper, and the peak intensities get higher. Since we do not know the true gains, how to define the variances of gains from ALMA observation information is an open problem. The condition of the atmosphere would be directly related to the variances, and further study is required. 

The history of CLEAN and traditional self-calibration is long. There are a lot of technical implementations and it is not easy to replace all of them at once. It is necessary to evaluate the performance of the proposed framework using various emission models under different observational conditions. The method must be tested with other ALMA data, and the results must be compared closely with traditional CLEAN methods. This is one of our ongoing projects.

We believe the mathematical framework shown in this paper is flexible and will provide a good platform for further development. Some simple extensions include but not limited to multiband imaging, polarization imaging, and spectral line imaging.

Finally, we mention our software packages. The RML imaging algorithm has already been made available to the public as a PRIISM package, and we will soon provide the self-calibration part.

\begin{ack}
We thank Kazunori Akiyama, Mareki Honma, Mirai Tanaka, Shota Takahashi, and Ryohei Kawabe for their valuable discussion.
This work was financially supported in part by JSPS KAKENHI Grant Numbers JP20H01951 and JP23K20035. 
M.Y. acknowledges supports from the National Science and Technology Council (NSTC) of Taiwan with grant NSTC 112-2124-M-001-014 and NSTC 113-2124-M-001-008.
This paper makes use of the following ALMA data:
ADS/JAO.ALMA \#2015.100425.S, \#2011.0.00015.SV, and \#2011.0.00016.SV.
ALMA is a partnership of ESO (representing its member states), NSF (USA), and NINS (Japan), together with NRC (Canada), MOST and ASIAA (Taiwan), and KASI (Republic of Korea), in cooperation with the Republic of Chile. The Joint ALMA Observatory is operated by ESO, AUI/NRAO, and NAOJ.
\end{ack}

\appendix

\section{Solving Self-Calibration Problem\label{sec:details of algorithm}}

We extended the optimization algorithm in \cite{2017.Repetti.mnras} to solve equation\,(\ref{eq:selfcal_new}). We first duplicate the gains $\bm{g}$ to $\bm{g}$ and $\bm{h}$ and extend $L_{\Obsv}(\bm{x},\bm{g})$ and $S_{\mu_1,\mu_2}(\bm{g})$ as follows,
\begin{eqnarray}
\label{eq:extend L and S}
    &&\mathcal{L}_{\Obsv}(\bm{x},\bm{g},\bm{h})\\
    \nonumber
    &&\hspace{1em}
    =
    \sum_{k} 
    \frac{1}{4\sigma_k^2}
    \Bigl[
    \bigl| \tilde{v}_k g_{\alpha l}h_{\beta l}^\ast - \mathcal{F}_k(\bm{x})\bigr|^2
    + \bigl| \tilde{v}_k h_{\alpha l}g_{\beta l}^\ast - \mathcal{F}_k(\bm{x})\bigr|^2
    \Bigr]\\
    &&
    \mathcal{S}_{\mu_1,\mu_2}(\bm{g},\bm{h})\\
    \nonumber 
    &&\hspace{1em}
    =
    \frac{\mu_1}{2}
    \sum_{\alpha,l}
    w_{\alpha l}
    \Bigl[
    \bigl| g_{\alpha l}-h_{\alpha\,l-1}\bigr|^2
    +
    \bigl| h_{\alpha l}-g_{\alpha\,l-1}\bigr|^2
    \Bigr]
    \\
    \nonumber
    &&
    \hspace{2em}
    +\frac{\mu_2}{2}
    \sum_{\alpha,l}
    w_{\alpha l}
    \Bigl[
    \bigl( |g_{\alpha l}|-|h_{\alpha\,l-1}|\bigr)^2
    +
    \bigl( |h_{\alpha l}|-|g_{\alpha\,l-1}|\bigr)^2
    \Bigr].
\end{eqnarray}
Note that by setting $\bm{g}=\bm{h}$, $\mathcal{L}_{\Obsv}(\bm{x},\bm{g},\bm{g})=L_{\Obsv}(\bm{x},\bm{g})$ and $\mathcal{S}_{\mu_1,\mu_2}(\bm{g},\bm{g})=S_{\mu_1,\mu_2}(\bm{g})$ hold. Instead of solving equation\,(\ref{eq:selfcal_new}), we consider the following optimization problem.
\begin{eqnarray}
    \label{eq:selfcal_solve1}
    &&
    \hat{\bm{g}}
    =
    \argmin{\bm{g}}
    \Bigl[
    \mathcal{L}_{\Obsv}(\bm{x},\bm{g},\hat{\bm{h}}) + \mathcal{S}_{\mu_1,\mu_2}(\bm{g},\hat{\bm{h}})\\
    \nonumber
    &&\hspace{10em}
    +\frac{\rho}{2} \sum_{\alpha,l} \bigl| g_{\alpha l}-\hat{h}_{\alpha l}\bigr|^2
    \Bigr],\\
    \nonumber
    &&\mbox{subject to}
    \hspace{1em}
    \sum_{l} |g_{\alpha l}| = N_\alpha.
\end{eqnarray}
We now propose the following algorithm.    
\begin{enumerate}
    \item Initialize $\hat{\bm{h}} = \bm{1}$ and set $\rho$ to a small positive value.
    \item Update $\bm{g}$ by solving equation\,(\ref{eq:selfcal_solve1}).
    \item Set $\bm{h}=\hat{\bm{g}}$ and go to Step 2.
    \item Repeat Steps 2 and 3 until convergence.
    \item Increase $\rho$ and repeat Steps 2, 3, and 4. Exit when $\bm{g}$ and $\bm{h}$ are sufficiently close. 
\end{enumerate}

Starting with a small $\rho$, solve the above equation with respect to $h_{\alpha l}$ and $g_{\alpha l}$ alternatively. By increasing $\rho$, $g_{\alpha l}$ and $h_{\alpha l}$ become almost identical. The optimization problem of equation\,(\ref{eq:selfcal_solve1}) is quadratic with respect to $\hat{\bm{g}}$ and is solved in a closed form. The solution of equation\,(\ref{eq:selfcal_solve1}) is given as follows,
\begin{equation}
\label{eq:solution_form}
    \hat{g}_{\alpha l}
    =
    {r}_{\alpha l}
    \frac{b_{\alpha l}}{|b_{\alpha l}|}.
\end{equation}
where
\begin{eqnarray}
\label{eq:coefficients}
    &&
    r_{\alpha l}
    =
    \frac{1}{a_{\alpha l}}(|b_{\alpha l}| + c_{\alpha l} - \eta_\alpha),
    \\\nonumber
    &&
    a_{\alpha l} = 
    \sum_{k \in \mathcal{N}(\alpha l)}\frac{|\tilde{v}_k|^2 |h_{\beta l}|^2}{\sigma_k^2}
    + \rho + (\mu_1+\mu_2)(w_{\alpha l} + w_{\alpha\, l+1})\\\nonumber
    &&
    b_{\alpha l} = 
    \sum_{k \in \mathcal{N}(\alpha l)}\frac{y_k \tilde{v}_k^\ast h_{\beta l}}{\sigma_k^2}
    {+}\rho h_{\alpha l}
    {+}
    {\mu_1}
    (
    w_{\alpha l}h_{\alpha\, l-1}
    {+}
    w_{\alpha\, {l+1}}h_{\alpha\, l+1}
    )
    \\\nonumber
    &&
    c_{\alpha l} = 
    {\mu_2}
    \Bigl(
    w_{\alpha l}|h_{\alpha\, l-1}|
    +
    w_{\alpha\,{l+1}}|h_{\alpha\, l+1}|
    \Bigr),\\\nonumber
    &&
    \eta_\alpha = 
    \Bigl(
        \sum_{l}
        \frac{|b_{\alpha l}|+c_{\alpha l}}{a_{\alpha l}} -N_\alpha
    \Bigr)
    {\big/}
    \sum_{l}
    \frac{1}{a_{\alpha l}}.
\end{eqnarray}
Here, $\mathcal{N}(\alpha l)$ is the set of the visibility indexes in which $g_{\alpha l}$ is included. Finally, if an $r_{\alpha l}$ becomes negative, we set corresponding $g_{\alpha l}$ to 0. This corresponds to removing some data points. We note that this did not happen in the three imaging experiments of the paper.

\section{Estimated gains\label{sec:station gains}}

\begin{figure*}
    \begin{center}
        \includegraphics[width=\textwidth]{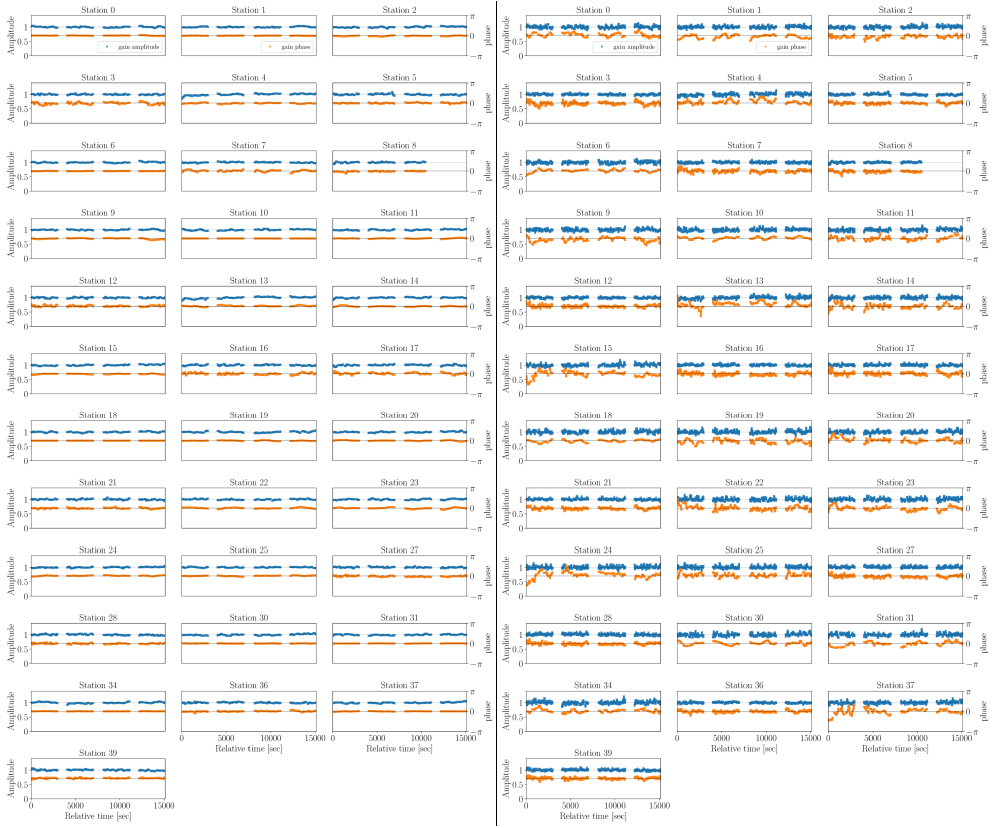}
    \end{center}
    \caption{The estimated gains of each station for HL\,Tau. Left and right show the estimated gains with \EstS and \EstL, respectively. Alt text: On the left and right sides: 34 scatter plots of the phases and amplitudes as a function of time. The plots on the left side have smaller variances than those on the right side.}
    \label{fig:HLTau station gains}
\end{figure*}

\begin{figure*}
    \begin{center}
        \includegraphics[width=\textwidth]{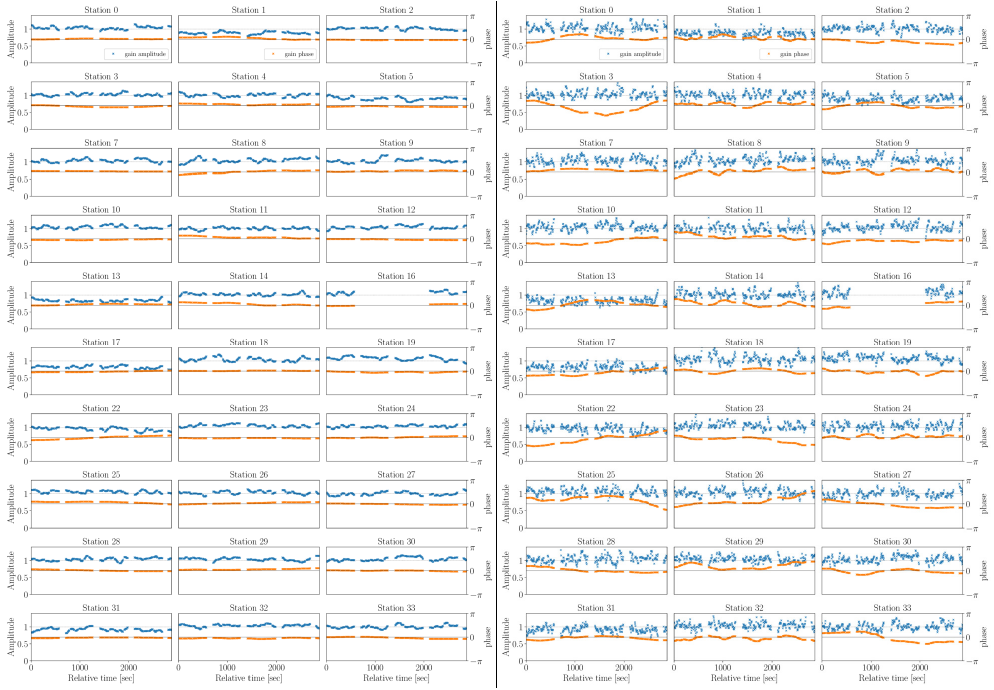}
    \end{center}
    \caption{The estimated gains of each station for SDP.\,81, part 1. Left and right show the estimated gains with \EstS and \EstL, respectively. Alt text: On the left and right sides: 30 scatter plots of the phases and amplitudes as a function of time. The plots on the left side have smaller variances than those on the right side.}
    \label{fig:SDP81 station gains 0}
\end{figure*}

\begin{figure*}
    \begin{center}
        \includegraphics[width=\textwidth]{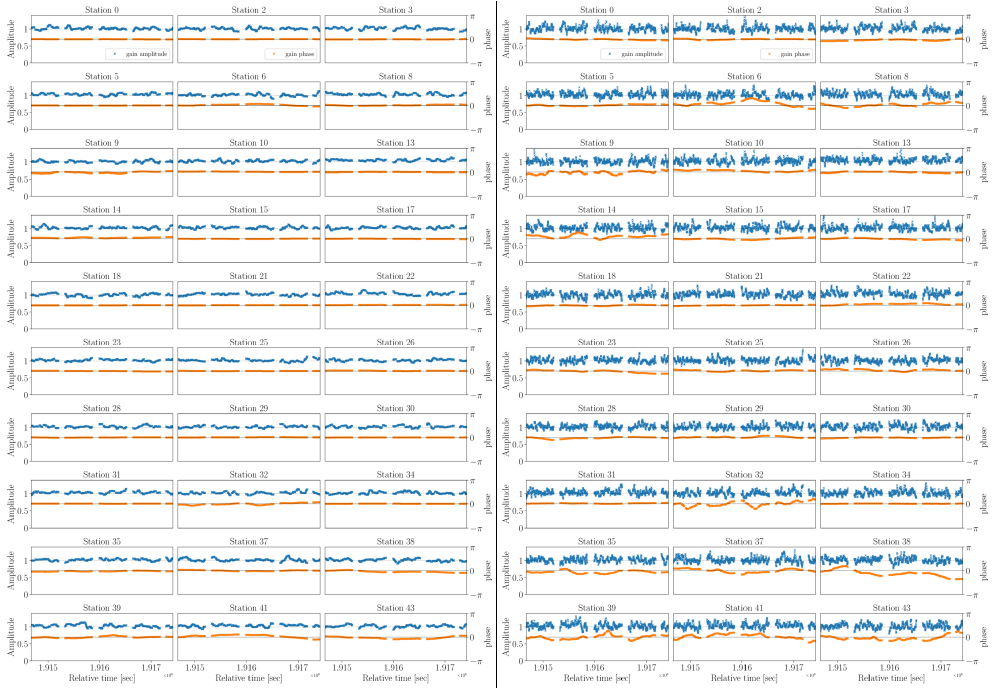}
    \end{center}
    \caption{The estimated gains of each station for SDP.\,81, part 6. Left and right show the estimated gains with \EstS and \EstL, respectively. Alt text: On the left and right sides: 30 scatter plots of the phases and amplitudes as a function of time. The plots on the left side have smaller variances than those on the right side.}
    \label{fig:SDP81 station gains 5}
\end{figure*}

\begin{figure*}
    \begin{center}
        \includegraphics[width=\textwidth]{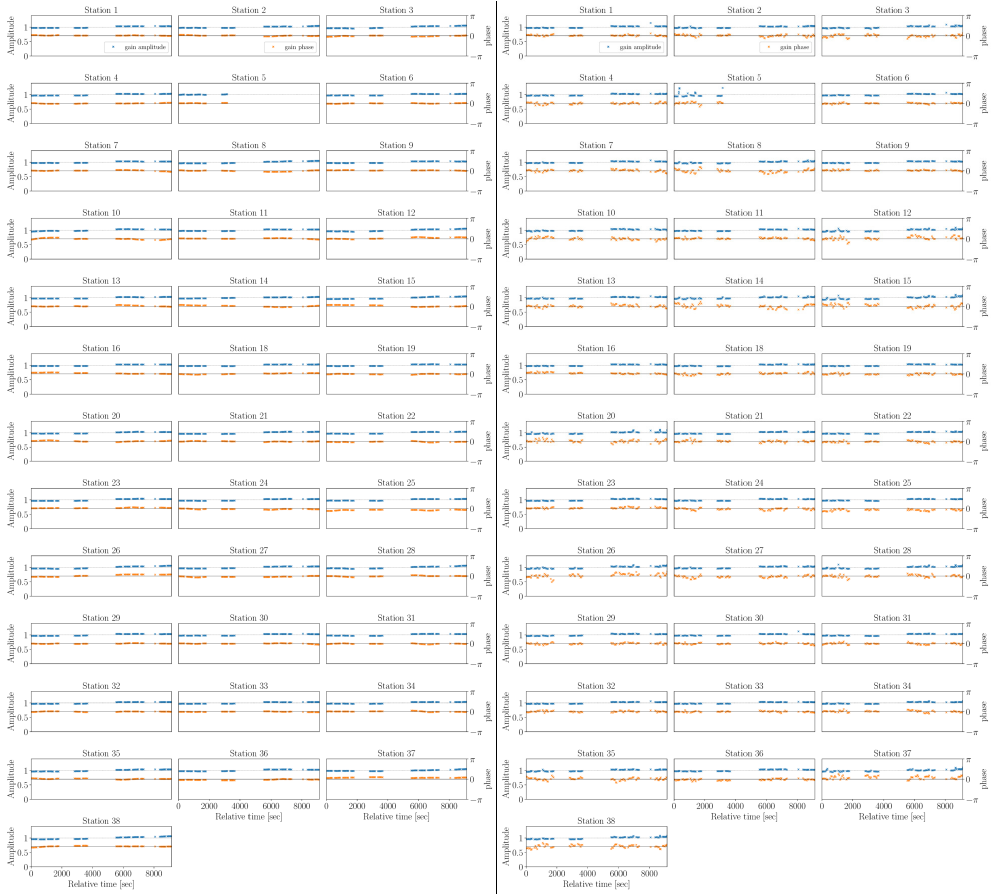}
    \end{center}
    \caption{The estimated gains of each station for HD\,142527. Left and right show the estimated gains with \EstS and \EstL, respectively. Alt text: On the left and right sides: 37 scatter plots of the phases and amplitudes as a function of time. The plots on the left side have smaller variances than those on the right side.}
    \label{fig:HD142527 station gains}
\end{figure*}

Finally, we show the estimated gains as time series. Figures\,\ref{fig:HLTau station gains}, \ref{fig:SDP81 station gains 0}$\sim$\ref{fig:SDP81 station gains 5}, and \ref{fig:HD142527 station gains} show the gains of HL\,Tau, SDP.\,81, nx HD\,142527, respectively. The data of SDP.\,81 is a concatenation of 6 different days, and each figure shows the gains of each day. The Left and right sides of each figure show the results with \EstS and \EstL, respectively. Each figure shows how the phases and the amplitudes of the estimated gains change over time. The estimated gains, especially the phases, change smoothly over time. 


\begin{figure*}
    \begin{center}
        \includegraphics[width=\textwidth]{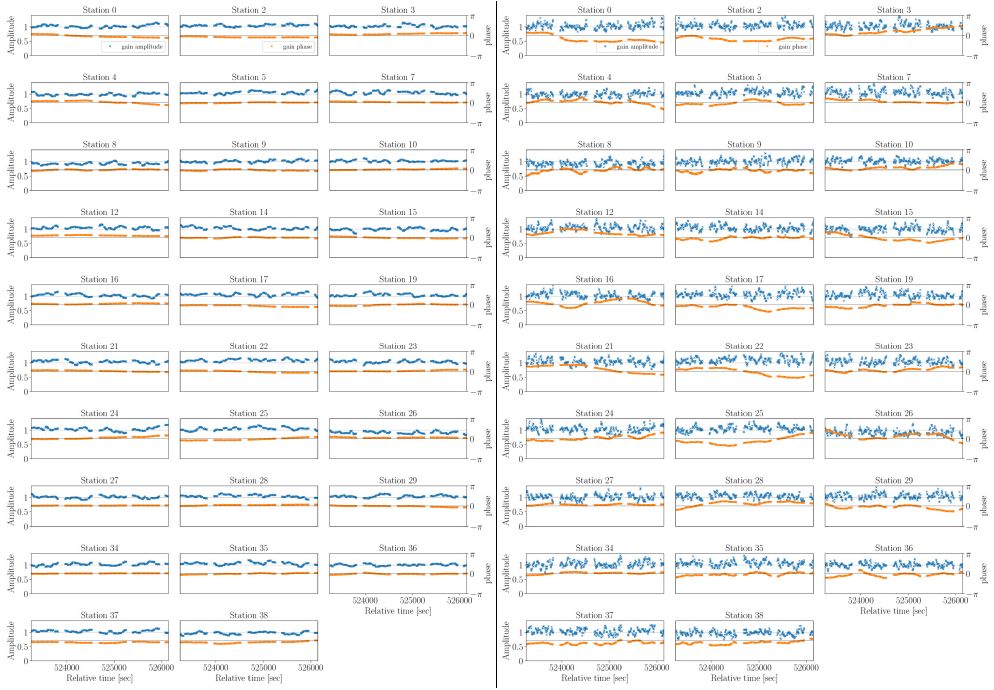}
    \end{center}
    \caption{The estimated gains of each station for SDP.\,81, part 2. Left and right show the estimated gains with \EstS and \EstL, respectively. Alt text: On the left and right sides: 29 scatter plots of the phases and amplitudes as a function of time. The plots on the left side have smaller variances than those on the right side.}
    \label{fig:SDP81 station gains 1}
\end{figure*}

\begin{figure*}
    \begin{center}
        \includegraphics[width=\textwidth]{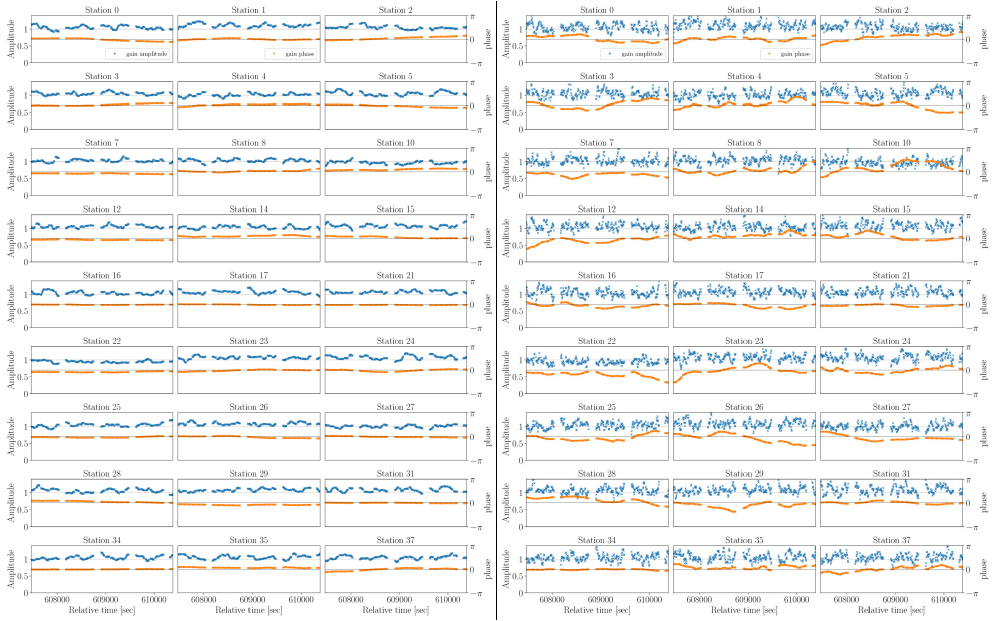}
    \end{center}
    \caption{The estimated gains of each station for SDP.\,81, part 3. Left and right show the estimated gains with \EstS and \EstL, respectively. Alt text: On the left and right sides: 27 scatter plots of the phases and amplitudes as a function of time. The plots on the left side have smaller variances than those on the right side.}
    \label{fig:SDP81 station gains 2}
\end{figure*}

\begin{figure*}
    \begin{center}
        \includegraphics[width=\textwidth]{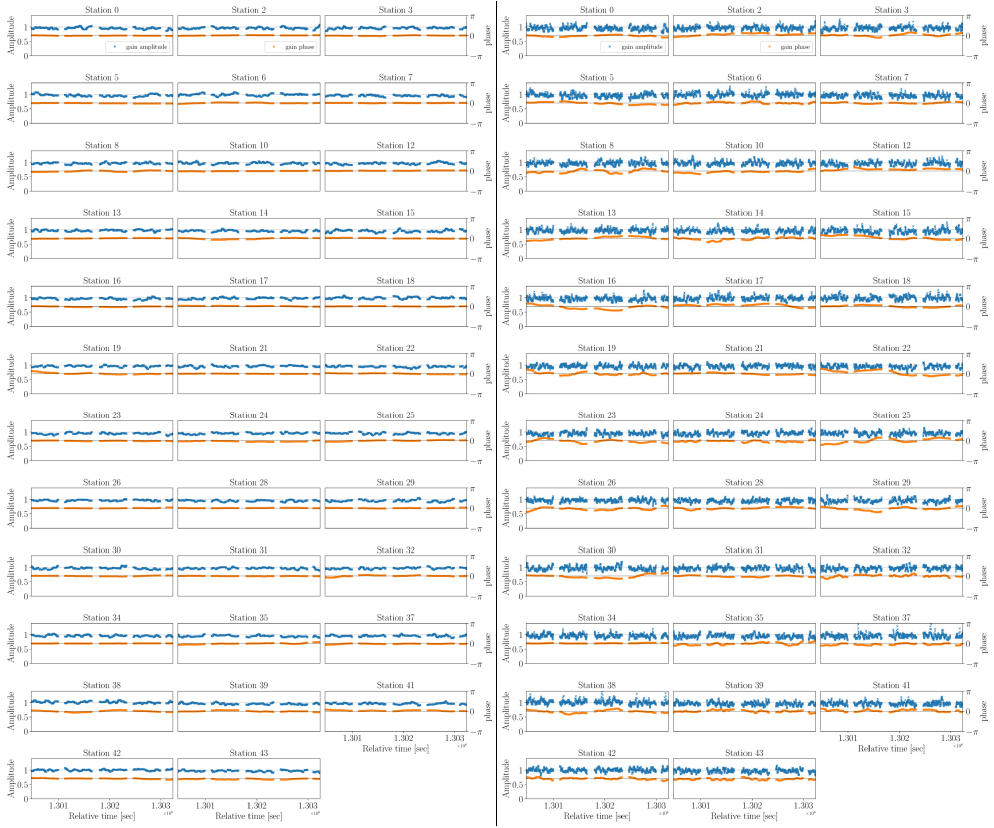}
    \end{center}
    \caption{The estimated gains of each station for SDP.\,81, part 4. Left and right show the estimated gains with \EstS and \EstL, respectively. Alt text: On the left and right sides: 35 scatter plots of the phases and amplitudes as a function of time. The plots on the left side have smaller variances than those on the right side.}
    \label{fig:SDP81 station gains 3}
\end{figure*}

\begin{figure*}
    \begin{center}
        \includegraphics[width=\textwidth]{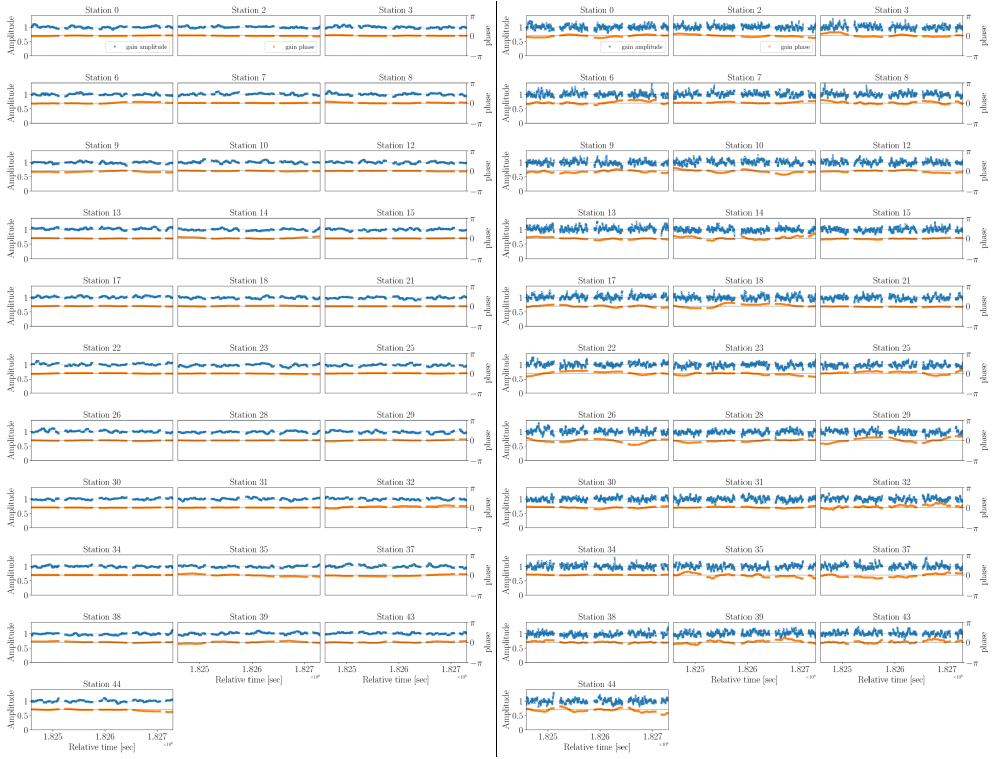}
    \end{center}
    \caption{The estimated gains of each station for SDP.\,81, part 5. Left and right show the estimated gains with \EstS and \EstL, respectively. Alt text: On the left and right sides: 31 scatter plots of the phases and amplitudes as a function of time. The plots on the left side have smaller variances than those on the right side.}
    \label{fig:SDP81 station gains 4}
\end{figure*}

\end{document}